# Hydrogen plasma exposure of In/ITO bilayers as an effective way for dispersing In nanoparticles

Zheng Fan[1], Jean-Luc Maurice[1], Stéphane Guilet[2], Edmond Cambril[2], Xavier Lafosse[2], Laurent Couraud[2], Kamel Merghem[2], Sophie Bouchoule[2], Pere Roca i Cabarrocas[1]

[1]LPICM, CNRS, Ecole Polytechnique, Université Paris-Saclay, 91128 Palaiseau, France

[2]C2N, CNRS, Université Paris-Saclay, Route de Nozay, 91460 Marcoussis, France

**Abstract**: We address the production of indium nanoparticles (In NPs) from In thin films thermally evaporated on both *c*-Si substrates and sputtered indium tin oxide (ITO) as well as from sputtered ITO thin films, exposed to a hydrogen ($H_2$) plasma. On the one hand, we show that evaporated In thin films grow in Volmer-Weber (VW) mode; $H_2$ plasma reduces their surface oxide and substrate annealing reshapes them from flat islands into spheres, without any remarkable surface migration or coalescence. On the other hand, we studied the In NPs formation on the ITO thin films and on In/ITO bilayer structures, by varying the $H_2$ plasma exposure time and the substrate temperature. This led us to postulate that the main role of $H_2$ plasma is to release In atoms from ITO surface. At low substrate temperature (100 °C), In NPs grow on ITO surface via a solid phase VW mode, similar to evaporated In thin films, while at 300 °C, small In droplets preferentially nucleate along the ITO grain boundaries where ITO reduction rate and atomic diffusion coefficient are higher compared with the ITO grain surface. As the droplets grow larger and connect with each other, larger ones (1-2 μm microns) are suddenly formed based on a liquid phase growth-connection-coalescence process. This phenomenon is even stronger in the case of In/ITO bilayer where the large In drops resulting from the evaporated In connect with the smaller NPs resulting from ITO reduction and rapidly merge into very large NPs (15 μm).

**Introduction**

Semiconductor nanowires are gaining increasing interest in diverse applications such as electronics, photonics, biosensors, photovoltaics, etc. [1, 2] For bottom-up growth methods, metal droplets are commonly employed as catalysts for the nanowires growth [3]. Gold (Au) was the first metal chosen for silicon nanowire growth [4], and since then it has been widely used for various semiconductor nanowires such as silicon [5], germanium [6] III-V materials [7]. However, Au contaminates semiconductors and inevitably introduces deep level recombination centers [8], resulting in their electrical properties degradation. Indium is an alternative candidate for nanowire growth and various methods of preparing In catalysts have been reported. As a foreign element, In droplets can catalyse the vapour-solid-liquid (VLS) growth of silicon [9] and germanium [10] nanowires, or even migrate on hydrogenated amorphous silicon (*a*-Si:H) and lead the growth of in-plane silicon nanowires obtained via a solid-liquid-solid process [11]. Evaporation from a pure source is a simple and direct way to deposit In on substrates. On the other hand, In-based III-V nanowires such as InP [12] and InAs [13] can be self-catalysed by In droplets, which are directly deposited on substrates by dissociating precursors such as trimethylindium (TMI) [14].

Besides the deposition methods, it has been reported that In NPs can be generated *in-situ* from the reduction of ITO (or $In_2O_3$) thin films by a $H_2$ plasma treatment, from which silicon nanowires were successfully grown [15, 16, 17, 18, 19]. Moreover, based on this novel technology, silicon nanowire-based solar cells have been demonstrated [20]. As a type of transparent conductive oxide (TCO), it is well known that $H_2$ plasma reduces ITO thin films, producing and yielding In-rich surfaces [21], so that electrical and optical properties of the thin films can be modified [22, 23, 24]. However, controlling the NPs density and size distribution (e.g. NPs with diameters of 50~100 nm and spacing of 500 nm to 1 μm) is still a big challenge for the optimisation of silicon nanowire-based solar cells [25, 26]. Indeed, patterning the metal NPs by e-beam or nanoimprint lithography is a straight forward way for realising well-organised nanowire arrays [27]. However, we are pursuing a low cost route which would be preferred by the industry. With this objective, we carried out a systematic study on the evolution of evaporated In and sputtered ITO, from thin films to NPs under $H_2$ plasma exposure.

**Experiments**

In order to study the effects of $H_2$ plasma and substrate temperature on i) In droplets migration on *c*-Si and ITO substrates, ii) ITO reduction, iii) In atoms surface diffusion and nucleation on ITO surface, we prepared three types of samples: type (1), thermally evaporated In thin films on n-type (100) *c*-Si substrate (with native oxide), in nominal thicknesses of 5, 50, 100, 200 and 500 nm; type (2), 200 nm thick RF-magnetron sputtered ITO thin films on *c*-Si, at room temperature (RT) and 350 °C; type (3), 50 nm In pads (200x200 μm) evaporated on 200 nm RT-sputtered ITO thin films, defined by optical lithography. A 5 nm evaporated In thin film was also deposited on a $Si_3N_4$ membrane for transmission electron microscopy (TEM) analysis. The as-sputtered ITO thin films were characterised by X-ray diffraction (XRD). The samples were loaded in a RF-PECVD system at a nominal substrate temperature ($T_{sub}$) of 150 °C, and pumped down to a vacuum level of $5x10^{-6}$ mbar. Afterwards, the samples were treated by a standard $H_2$ plasma (100 sccm $H_2$, 400 mTorr, RF power of 5 watts) at different $T_{sub}$ for various durations. Note that the substrate temperature was stabilised in $H_2$ atmosphere (100 sccm, 1.2 Torr) for 15 minutes prior to the $H_2$ plasma treatment. After cooling down to $T_{sub}$ of 150 °C, the samples were unloaded and transferred to a scanning electron microscope (SEM) for observation. Fig. 1 lists all the tests carried out in this study:

Test I: (samples of type 1) evaporated In thin films were treated by a standard $H_2$ plasma at 300 °C for 5 minutes.

Test II: (samples of type 2) RT and 350 °C sputtered ITO thin films were treated by a standard $H_2$ plasma at 100 °C for 1, 5, 30, 60 minutes.

Test III: (samples of type 2) RT and 350 °C sputtered ITO thin films were treated by a standard $H_2$ plasma at 300 °C for 15, 30, 45, 60 seconds.

Test IV: (samples of type 3) 50 nm evaporated In/RT-sputtered ITO bilayer were treated by 5 % diluted hydrogen chloride (HCL) in deionised water (by volume) etching for 3 seconds and then annealed at 300 °C in Ar/$H_2$ atmosphere for 5 minutes.

Test V: (samples of Type 3) 10 or 50 nm evaporated In/RT-sputtered ITO bilayer were treated by a standard $H_2$ plasma at 300 °C for 5 minutes.

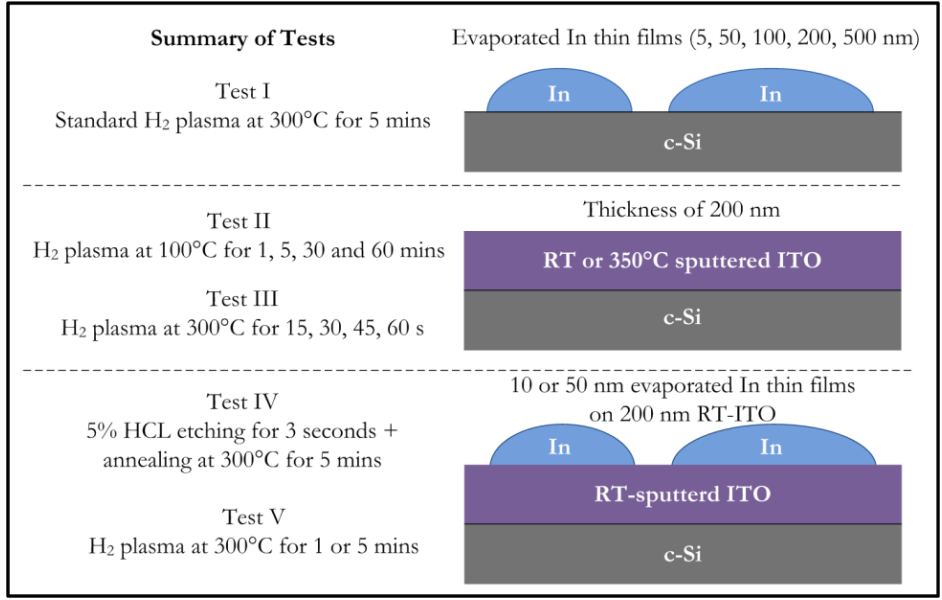

Fig. 1. Schematic representation of the various tests carried out on the different types of samples.

**Results and discussion**

### Evaporated indium thin films on *c*-Si substrate

Evaporation is the simplest way for preparing In catalyst. In Test I, we study the In thin film growth (by varying the thin film thicknesses from 5 to 500 nm), and the evolution of the thin films after $H_2$ plasma.

Indium is easily self-passivated in ambient atmosphere, forming surface indium oxide [28]. In order to achieve efficient In/semiconductor precursor contact and interaction, pre-treatment such as introducing hydrogen (H) radicals [9, 29] is needed to remove the native surface oxide. Fig. 2 (a) shows the TEM image of an as-evaporated In NP (evaporated In thin films have a morphology of discontinuous NPs or islands, as discussed below). Direct evidence for the presence of an In oxide shell can be obtained by nano-scale electron energy loss spescopy (EELS) analysis. The intensity of the edge is determined by the amount of the corresponding material. Fig. 2 (b, c) show the EELS spectra of the center and the surface of the NP, respectively. As the intensity ratio of O-K edge over In-$M_{4,5}$ edge is larger on the surface, we confirm that the NP surface was oxidised.

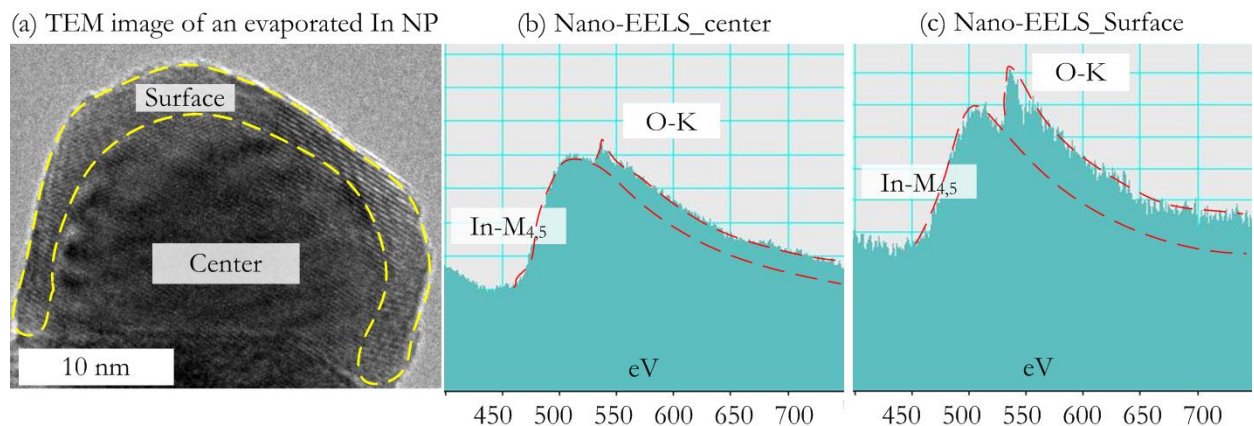

Fig. 2 TEM study of as-evaporated In: (a) TEM image of an evaporated In NP; (b, c) nano-scale EELS spectra taken from the center and from the surface of the NP, respectively, showing different intensity ratio of O-K edge over In-$M_{4,5}$ edge, which demonstrate the presence of a surface oxide.

Figure 3 (a-1 to e-1) shows the morphology of evaporated In thin films on *c*-Si with increasing nominal thicknesses (5, 50, 100, 200 and 500 nm, respectively), which obeys the Volmer-Weber (VW) growth mode (or island growth mode) [30]. The evaporated In thin films evolved from small NPs to discontinuous flat islands and finally the growing islands overlapped and coalesced into continuous films. Image (a-1) shows the dense NPs with narrow size distribution. Images (b-1, c-1 and d-1) show the growth of discontinuous islands, which are flat islands as seen in image (b-4). Images (b-3, c-3 and d-3) imply that during the large island growth, new small NPs were growing between them. Image (e-1 and e-3) shows the continuous film after island growth and coalescence.

After a standard $H_2$ plasma at 300 °C for 5 minutes (above the In melting point 157 °C [31]), we observed that the flat islands reshaped into spheres, as can be seen from images (b-2) to (d-2). An example of the spheres is seen in SEM image (b-6) in tilt angle of 75 °. Similarly, we suggest that the as-deposited small NPs also experienced this reshaping behaviour, which yields the enlarged spacing among them, as seen from image (a-1) to (a-2). Moreover, in images (b-5), (c-4) and (d-4), small NPs were still present between the large ones, which implies that no remarkable In droplets surface migration and coalescence occurred on *c*-Si substrates. However, after $H_2$ plasma exposure, continuous In films broke into largely separated spheres with no small ones among them (see images (e-2, e-4)), in comparison with the ones in images (c-2) and (d-2). Therefore, we conclude that the evolution of evaporated In thin films morphology (small NPs grow larger, coalesce into discontinuous islands and finally into continuous films) results in the evolution of In NPs after $H_2$ plasma, in a tendency from dense NPs of similar sizes (see images (a-2), (b-2)), to dispersed NPs in broad size distribution (see images (c-2), (d-2)) and finally to very large dispersed NPs with narrow size distribution (see images (e-2), (e-4)). This means that adjusting the In thin film morphology is a convenient way to control the density and size distribution of In NPs.

Based on the observations above, we can draw a scenario on the evolution of evaporated In thin films treated by $H_2$ plasma at 300 °C: $H_2$ plasma reduces the surface oxide of In thin films, allows In melting and wetting on *c*-Si substrate during substrate heating without any remarkable migration and coalescence. After cooling down the In droplets dewet and solidify into spheres. This reshaping behaviour of solid In from flat islands into spheres is supposed to be due to the minimization of Gibbs free energy as a sphere has the lowest surface area to volume ratio. A schematic representation is illustrated in Fig. 9 (a).

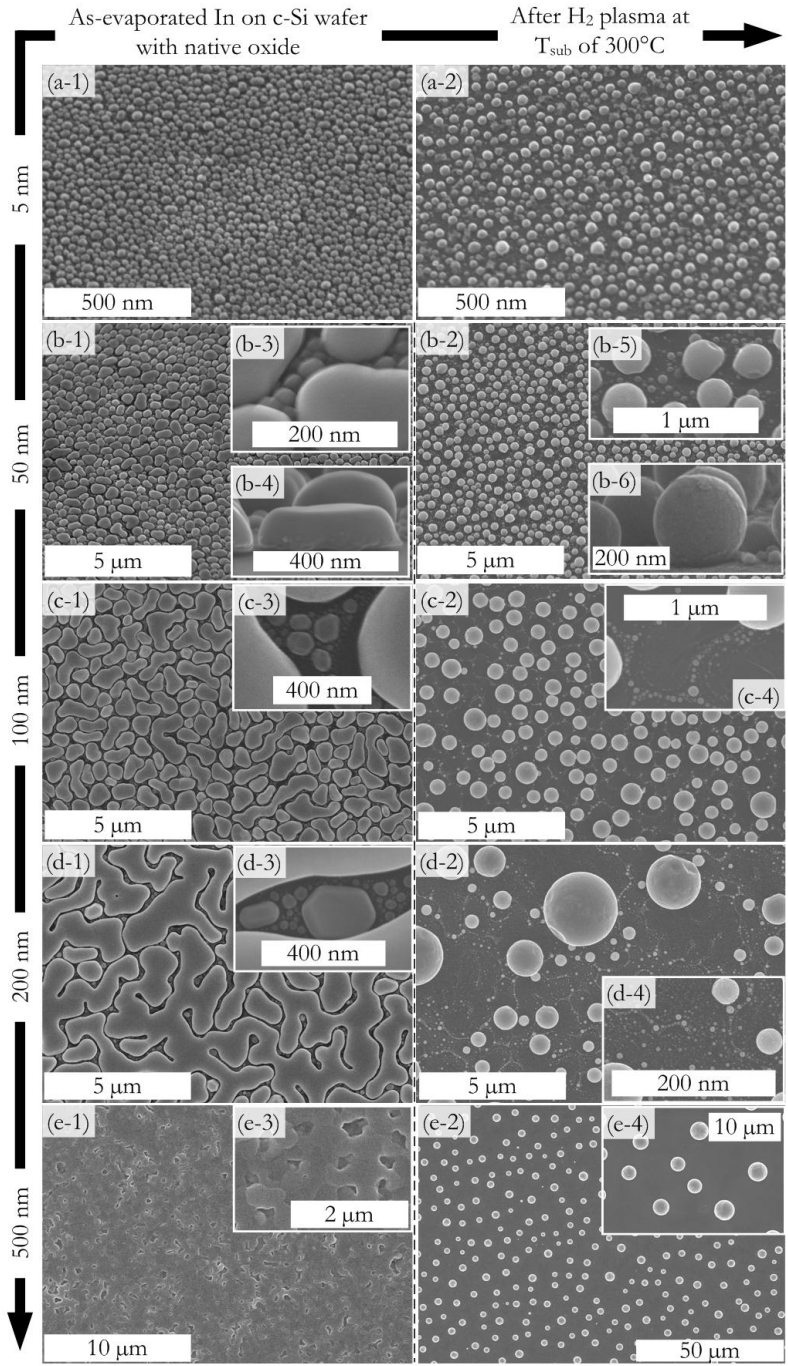

Fig. 3. Results of Test I: SEM images of evaporated In thin films with increasing thicknesses (from 5 to 500 nm) treated by a $H_2$ plasma at 300 °C for 5 minutes.

### ITO thin films sputtered at RT and 350 °C

As introduced above, the type 2 samples consist of ITO sputtered on *c*-Si substrates at two temperatures (RT and 350 °C). Prior to the $H_2$ plasma treatment, we first characterised the ITO properties. It is well known that the properties of ITO thin films are determined by their preparation conditions [32]. XRD patterns show that RT-sputtered ITO thin films are amorphous while the ones sputtered at 350 °C are polycrystalline, as shown in Fig. 4. Peaks corresponding to the (211), (222), (400), (440) orientations are marked, respectively. As shown below, the structure (amorphous or polycrystalline) of ITO has a strong impact on its reduction against $H_2$ plasma.

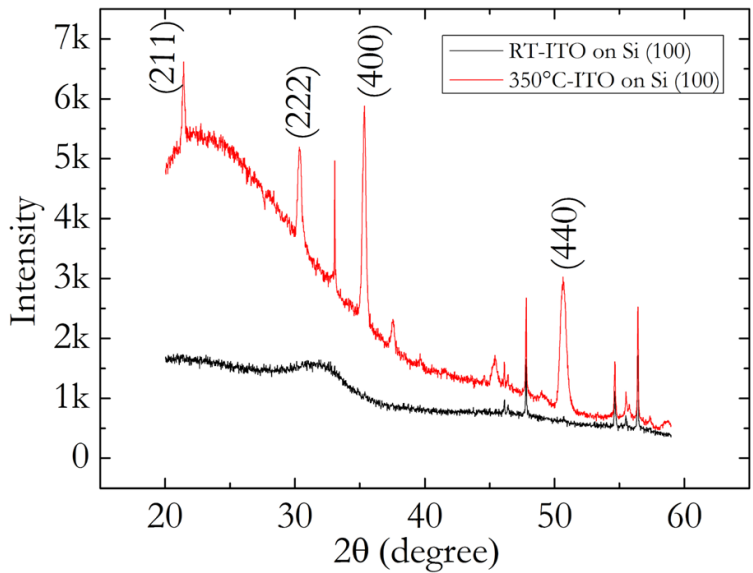

Fig. 4. XRD patterns of ITO thin films sputtered at RT and 350 °C, showing that the ITO thin films sputtered at RT are amorphous while the ones sputtered at 350 °C are polycrystalline.

The first series of $H_2$ plasma treatments was performed at 100 °C (i.e. Test II, see Fig. 1). Fig. 5 (a-1) shows that small NPs with narrow size distribution were formed on RT-sputtered ITO after 1 minute $H_2$ plasma. As the plasma duration increases, the small NPs grew larger, as shown in images (b-1) and (c-1), corresponding to 5 and 30 minutes of $H_2$ plasma, respectively. Note that in fact the NPs were InSn alloy with the mass ratio of In/Sn around 9:1 [33]. For simplicity, we will consider them as In NP. The inset images in (b-1) and (c-1) show that the NP size distributions were broadened with increasing $H_2$ plasma exposure time. For the moment, it remains unclear why there were some regions of the ITO surface with much fewer and smaller NPs, as marked by the yellow circles in images (a-1) and (b-1). However, after 30 minutes of $H_2$ plasma treatment, such kind of regions disappeared. As the $H_2$ plasma treatment continued to 60 minutes, we observed that the NPs did not grow larger any more, as shown in image (d-1), in comparison with the ones obtained by 30 minutes exposure. We suggest that at initial stage the evolution of In NPs is similar with the growth of evaporated In thin films, as shown in Fig. 3.

Therefore, we postulate that at 100 °C which is well below the In melting point, In NPs formation on ITO surface obeys the VW growth mode, where $H_2$ plasma reduces the ITO surface, releases In atoms on it, followed by In atoms surface diffusion, nucleation and growth into spheres rather than islands. However, in contrast with evaporation process, once the ITO surface is almost covered by dense In NPs, no more In atoms can be released, which is probably due to the fact the released In atoms form a thin layer which acts as barrier against the $H_2$ plasma/ITO interaction.

A similar tendency was observed for 350 °C-sputtered ITO, however the whole process was retarded. Fig. 5 (a-2) shows that 1 minute $H_2$ plasma can only transform the ITO surface to be rough and In-rich [21] without NPs formed. Furthermore, grain boundaries (GBs) were also observed on these polycrystalline ITO thin films. At 5 minutes, tiny NPs were formed (see image (b-2)) and as time went on they grew into larger ones at 30 minutes (see image (c-2)) and afterwards no more remarkable growth was observed at 60 minutes (see image (d-2)).

Compared with the ones formed on RT-sputtered ITO (see images (d-1) and (d-2)), the NPs were smaller with less broad size distribution. We suggest that this is due to the properties of ITO thin films. Obviously, the reduction rate by $H_2$ plasma is lower for polycrystalline 350 °C-sputtered ITO, compared with amorphous RT-ITO. This is consistent with the fact that the etching rate of polycrystalline ITO by HCL solution is much lower than the one of amorphous ITO [34]. As shown in Fig. 5, In atoms can only nucleate and grow into a large density of smaller NPs on polycrystalline 350 °C-ITO before they block the ITO reduction, compared with the larger ones on amorphous RT-ITO. This is probably due to the different diffusion coefficient on crystalline and amorphous surface. It is known that vacancy mechanism dominates the atomic diffusion in crystals, while besides vacancies, the random amorphous matrix can also mediate the atomic diffusion via interstitials [35]. Moreover, structure relaxation by thermal annealing efficiently decreases the atomic diffusion [36]. Therefore, we suggest that In atomic surface diffusion on 350 °C-ITO surface is lower than the one on RT-ITO surface. Even though the GBs with defects favour the etching [34] and atomic diffusion [37], we find that at 100 °C they had no significant effect on the In NP formation. Indeed, as shown in Fig. 5 (a-2 to d-2), the GBs quickly disappear under the In NPs. Fig. 9 (d) illustrates the evolution of ITO thin films by $H_2$ plasma treatment at 100 °C.

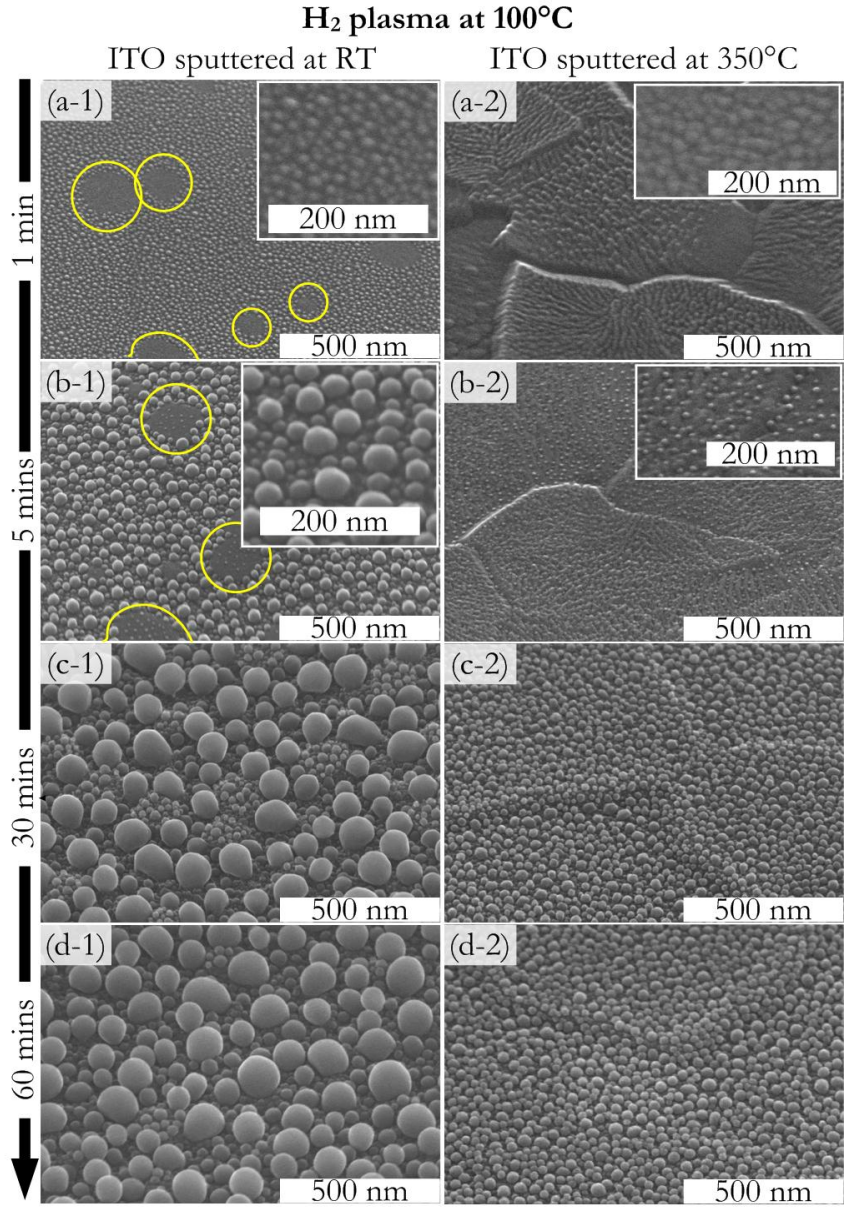

Fig. 5. Results of Test II: (a-d) SEM images of RT and 350 °C sputtered ITO thin films treated by a $H_2$ plasma at 100 °C for 1, 5, 30 and 60 minutes.

The second series of $H_2$ plasma treatments was performed at 300 °C (i.e. Test III, see Fig. 1). Fig. 6 (a-1) shows the RT-sputtered ITO surface treated by $H_2$ plasma for 15 seconds. Note that the reduction process is much faster at 300 °C, compared with that at 100 °C in Fig. 5. Interestingly, the ITO surface is separated into discontinuous regions, with small and dense NPs inside and relatively larger and dispersed ones (an example is marked by the yellow circle) along the paths. We suggest that the paths where the larger NPs are located represent the GBs of annealed RT-ITO thin films (due to the $T_{sub}$ stabilisation at 300 °C for 15 minutes, see the experiments section), as amorphous RT-ITO is known to start to crystallise at ~ 150 °C [38]. As $H_2$ plasma treatment went on (from 15, 30 to 45 seconds), the large NPs on the GBs grew larger while the In density decreased, as shown in Fig. 6 (b-1) and (c-1), respectively. A similar tendency was observed on 350 ℃-ITO surface, as shown in Fig. 6 (a-2) to (c-2). In particular, the larger NPs were indeed formed along the GBs and the fact that GBs remain clear at 300 ℃ compared to the ones at 100 ℃ (see Fig 5) suggest that In atoms diffuse faster along GBs at 300 ℃. However, after 60 seconds $H_2$ plasma, much larger NPs were surprisingly formed on both types of ITO, with lower density and much broader size distribution, as shown in images (d-1), (e-1) and (d-2), (e-2). The GBs did not trap these large NPs formation any more.

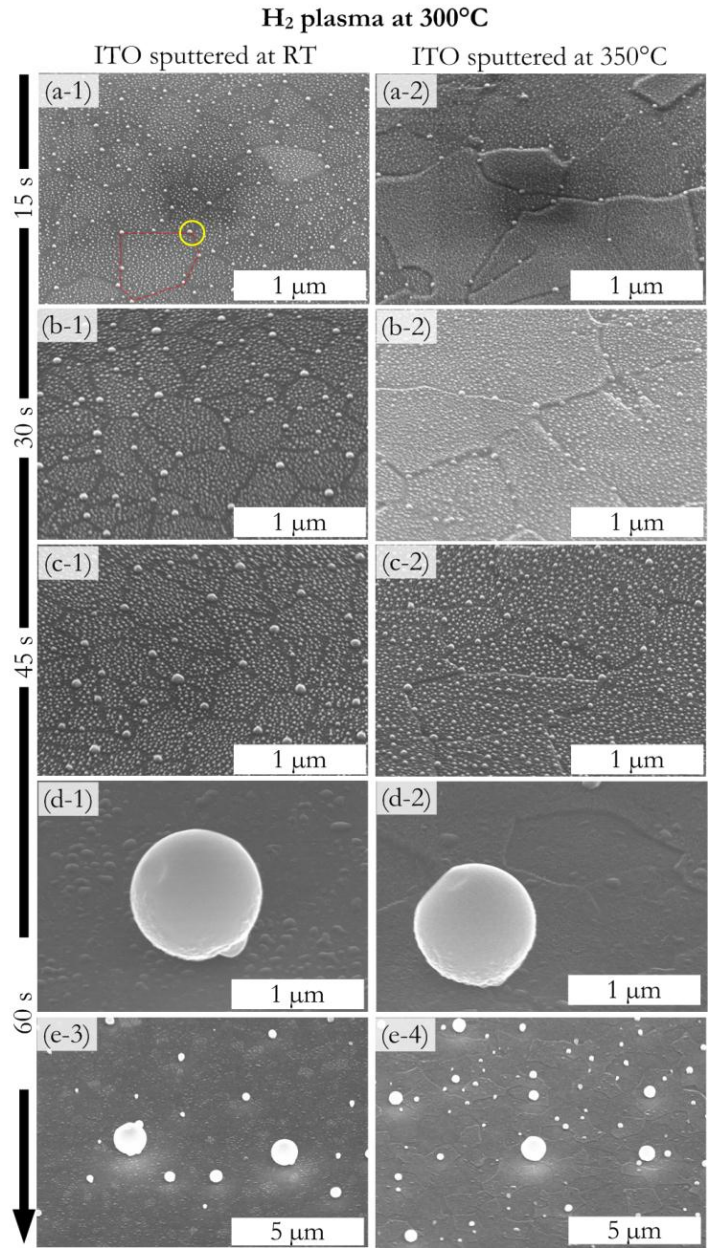

Fig. 6. Results of Test III: SEM images of RT and 350 ℃ sputtered ITO thin films treated by a $H_2$ plasma at 300 ℃ for 15, 30, 45 and 60 seconds, respectively.

Figure 7 shows the size and counts of In NPs formed by $H_2$ plasma from 15 to 60 seconds on both types of ITO, based on the SEM images Fig. 6 (a-1 to d-1) and (a-2 to d-2) with same magnification. From 15 to 45 seconds, the sizes of In NPs formed on GBs from both ITO thin films increased at very low rate (~0.7 nm/s) and the density also decreased linearly. Moreover, larger NPs can be formed on RT-ITO compared to 350 ℃-ITO. Similar with Test II ($T_{sub}$ = 100 ℃), we suggest that the annealed RT-ITO was more strongly reduced. From 45 to 60 seconds, the sizes increased drastically and the counts dropped down to 1.

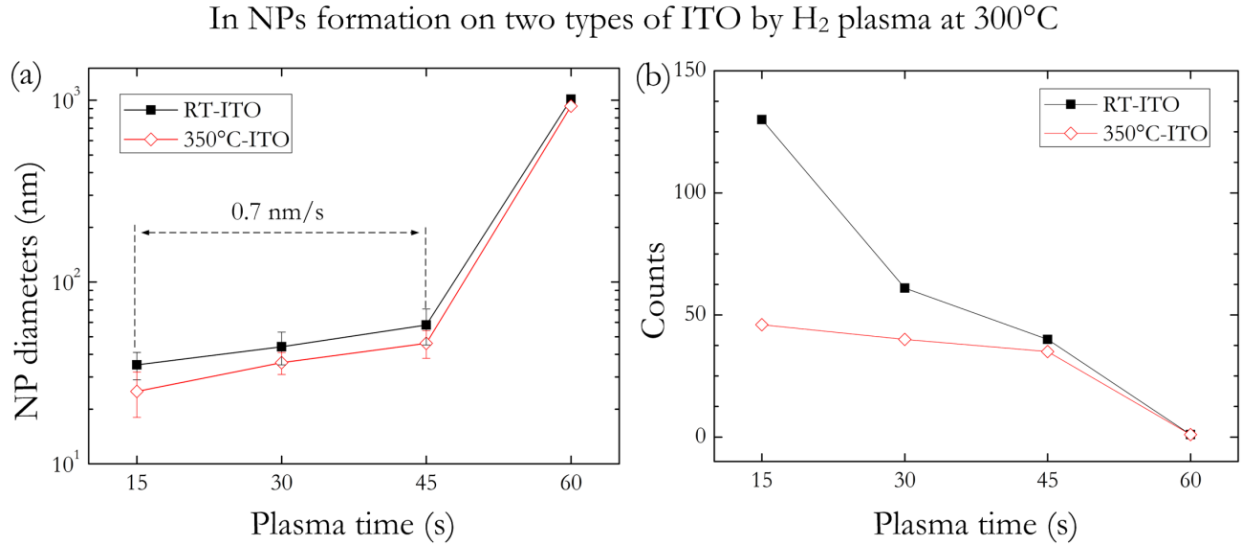

Fig. 7. Statistical analysis of the In NPs formed on GBs from ITO thin films sputtered at RT and 350 ℃, by a $H_2$ plasma at 300 ℃ for 15, 30, 45 and 60 seconds. (a) NP diameters as a function of $H_2$ plasma time. (b) Counts of NPs as a function of $H_2$ plasma time.

Test III shows that at a 300 ℃ (above the In melting point), the GBs are the preferential sites for forming larger In droplets at the initial stage (from 15 to 45 seconds). We suggest that at 300 ℃ it is clearly revealed that GBs with more defects facilitate the ITO reduction and In atomic diffusion, which have been discussed in Test II. Moreover, as GBs have higher surface energy than the grain surface [39], based on the GB wetting theory [40, 41], small droplets formed nearby the GBs probably spread towards the GBs and coalesce into larger ones. However, from 45 to 60 seconds, this GB effect is suddenly eliminated, with drastically growth of NPs in size and lowered density. Considering that the ITO reduction rate is a constant, In atoms cannot be released in a huge amount from 45 to 60 seconds. Therefore, we propose that during this period there must be a catastrophic coalescence of the In droplets. In order to understand this coalescence process, whether In droplets migrate on ITO surface or not, we designed Test IV and V.

### Evaporated indium/RT-sputtered ITO bilayer

Test I has demonstrated that In droplets do not migrate on *c*-Si surface. In order to verify the case on ITO surface, we designed Test IV. Fig. 8 (a) shows the zoom part of sample Type 3: nominal 50 nm In pads evaporated on RT-sputtered ITO, showing that In thin films also follow VW growth mode on ITO surface, similar to the case of In evaporated on *c*-Si (see Test I). Before annealing the sample at 300 °C, it is necessary to remove the surface oxide from the evaporated In. In order to discriminate the ITO reduction by $H_2$ plasma, we chose wet etching to remove the surface oxide. Fig. 8 (b) shows that the outermost shells of In islands were smoothly etched by 5% HCL for 3 seconds. Note that the ITO was also etched by HCL, however, unlike $H_2$ plasma, after etching the surface was still ITO. After HCL etching and sample drying by nitrogen flow, the sample was transferred rapidly to a thermal treatment furnace and annealed at 300 °C in Ar/$H_2$ atmosphere for 5 minutes. Fig. 8 (c) shows that the substrate heating turned the HCL-etched In islands into spheres, which is in agreement with Test I (see Fig. 2).

Test IV demonstrates that annealing at 300 °C is not sufficient to activate the In droplets surface migration on ITO surface. The question is whether it is the same in a $H_2$ plasma environment. In order to verify it, we carried out Test V: $H_2$ plasma treatment on the same evaporated In/sputtered ITO bilayer at 300 °C for 5 minutes. Quite large In particles (up to 15 μm) were formed on the In/ITO bilayer pads. In contrast, none of such particles was formed on ITO surface without In, as shown in Fig. 8 (d). Another important observation is the surface morphology of In/ITO bilayer after $H_2$ plasma. We selected the area nearby the 15 μm particle, as shown in Fig. 8 (e). We assume that the surface roughness was transferred from the evaporated In thin film morphology (see image (a)), which is also probably due to the blocking effect against $H_2$ plasma by the evaporated In islands (see Test II). This indicates that the evaporated In droplets coalesced during $H_2$ plasma and contributed to the large In particles formation. As the main role of $H_2$ plasma is to reduce ITO surface and release fresh In atoms, we suggest that the fresh In atoms promote the droplets coalescence: 1) they can contribute to the growth of evaporated In droplets, 2) or probably formed new ones. Both types of events eventually cause the neighbouring In droplets connecting with each other, so that a massive coalescence is realised, as illustrated in Fig. 9 (b). The final solidified NPs are composed by evaporated In (from vapour source) and released In atoms from ITO surface (from solid source).

Similarly, we can employ this growth-connection-coalescence mechanism to explain the In NPs formation on ITO thin films: the released In atoms from ITO surfaces diffuse and nucleate into tiny In droplets, they grow larger and coalesce after they are connected with each other, as illustrated in Fig. 9 (c). In contrast with In/ITO bilayer, there is no external supply of In on ITO surface before $H_2$ plasma treatment, so the sizes of In NPs are much smaller.

Moreover, we found that by inserting an ITO layer between evaporated In and *c*-Si substrates (i.e. In/ITO bilayer), it is possible to significantly disperse In NPs with relatively narrow size distribution by a $H_2$ plasma treatment, thanks to the growth-connection-coalescence process. An example is shown in Fig. 8 (f), where 10 nm In on RT-ITO bilayer was treated by 5 minute $H_2$ plasma at 300 °C. This is a remarkable improvement compared with NPs formation from evaporated In (see Fig. 3) or from ITO thin films (see Fig. 5 and 6).

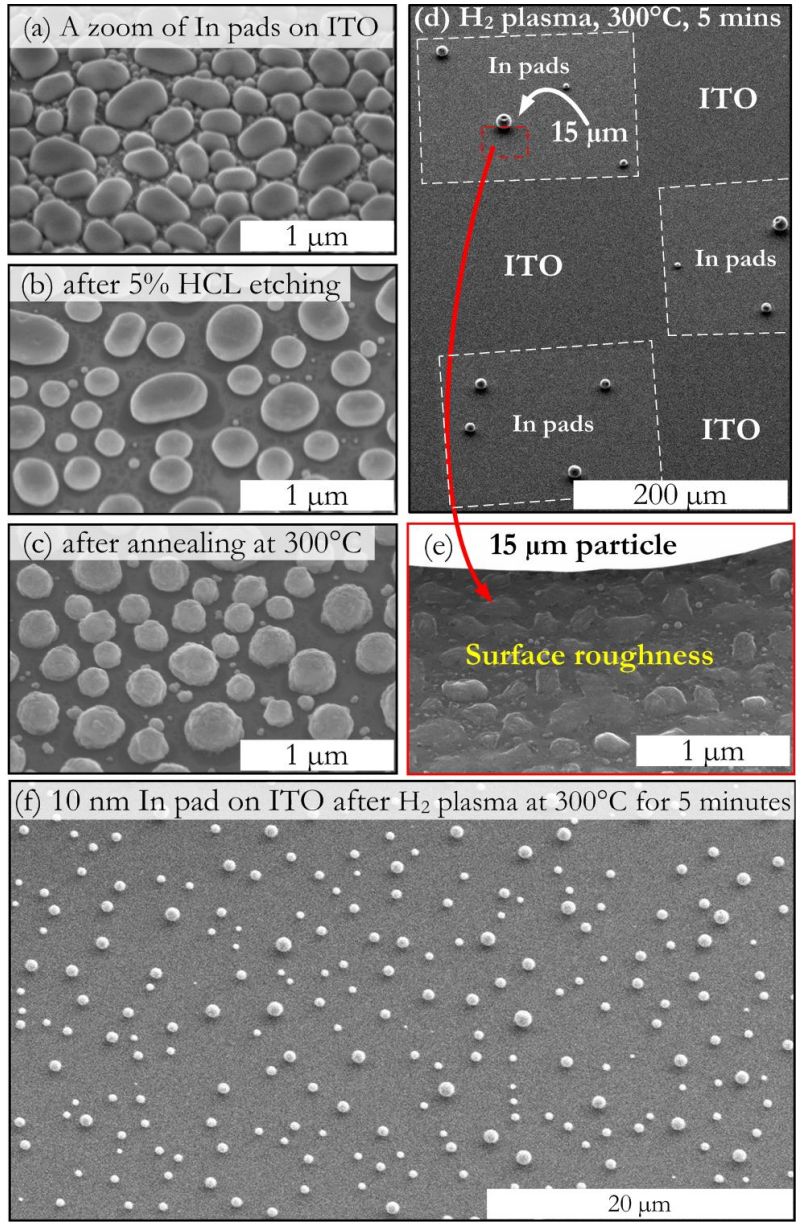

Fig. 8. Results of Test IV and V: (a) SEM image of nominal 50 nm In evaporated on RT-sputtered ITO thin film. Note that in fact this is a zoom of 200x200 µm In pads on ITO (i.e. samples of type 3). (b to c) SEM images of In/ITO bilayer etched by 5 % hydrogen chloride (HCL) (by volume) for 3 seconds and then annealed at 300 °C in Ar/$H_2$ atmosphere for 5 minutes. (d) SEM images of the same In pads on ITO exposed to a $H_2$ plasma at 300 °C for 5 minutes, showing that very large In particles were formed on the region of In/ITO bilayer, while none of such particles was formed on the ITO surface without evaporated In. (e) A zoom of the ITO surface around the large In particle after $H_2$ plasma, leaving a rough surface which reflects the positions of as-evaporated In islands. (f) 10 nm In on RT-ITO after $H_2$ plasma at 300 °C for 5 minute, resulting in dispersed In NPs with narrow size distribution compared with the ones formed on evaporated In (see Fig. 3) and ITO thin films (see Fig. 5 and 6).

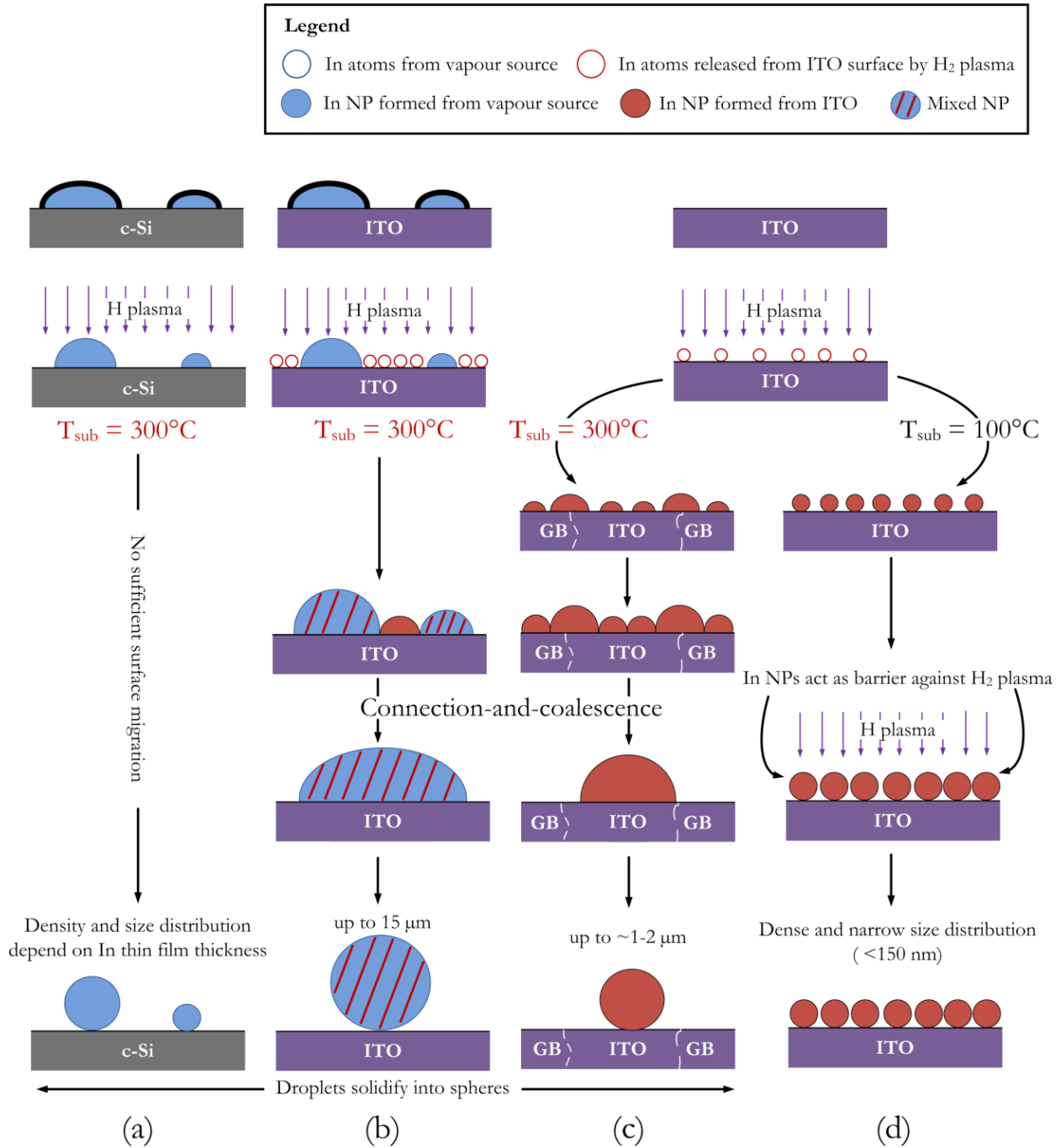

Fig. 9. Schematic representation of the evolution of evaporated In and sputtered ITO thin films exposed to a H$_2$ plasma. (a) In the case of evaporated In on *c*-Si, H$_2$ plasma at 300 °C reduces the surface oxide and allows the islands to reshape into spheres upon cooling. (b) In the case of In evaporated on ITO, H$_2$ plasma at 300 °C reduces both the surface oxide from the evaporated In and from the sputtered ITO thin film. The released In atoms from ITO contribute to the growth of evaporated In droplets or even nucleate into new ones, a massive coalescence takes place after the droplets connect each other which leads to large particles up to ~15 μm. (c) A similar process takes place on ITO surface where no evaporated In NPs are present. In particular, initially small NPs preferentially grow along the GBs where the ITO reduction rate and In atom diffusion

coefficient are higher than that on the grain surface. The final size of NPs is in the μm range. (d) Finally, when the sputtered ITO is exposed to the H$_2$ plasma at 100 °C, In atoms diffuse and nucleate in solid phase, NPs grow in Volmer-Weber mode in shape of spheres rather than island. The growing NPs finally block the H$_2$ plasma and the growth stops. More details are in the text.

**Summary and conclusion**

To summarise, we have studied the evolution of evaporated In and sputtered ITO thin films upon annealing and H$_2$ plasma exposure. Evaporated In thin films are formed in Volmer-Weber growth mode, where the source of In atoms comes from the vapour phase. At 300 °C, H$_2$ plasma reduces the surface oxide of In NPs and lets them wetting on *c*-Si substrates, without remarkable surface migration. The In NPs dewet and solidify into spheres due to the surface energy minimization. The density and size distribution of In NPs are strongly dependent on the morphology of as-evaporated In thin films. In contrast with evaporated In, ITO thin films act as solid phase source, from which In atoms are released by H$_2$ plasma. At low substrate temperature (100 °C), In NPs grow in the Volmer-Weber growth mode in solid phase. The reduction rate and In surface diffusivity depend on the crystallinity of ITO, so that In NPs grow smaller with a narrower size distribution on polycrystalline ITO sputtered at 350 °C, compared with the ones on amorphous RT-sputtered ITO. At high substrate temperature (300 °C), grain boundaries with more defects are preferential sites for the In droplets nucleation at initial stage, and a drastic growth in size and decrease in density take place via a growth-connection-coalescence mode in liquid phase.

These results provide guidelines for tailoring NPs size and density for the growth of silicon nanowires. On the one hand, it is possible to disperse In NPs by depositing thick In thin films (large islands), after H$_2$ plasma, tailoring them into right sizes by wet or dry etching. On the other hand, we propose to apply In/ITO bilayer structure to disperse In NPs with expected size distribution.

**Acknowledgments:** This work is partly supported by the French RENATECH network and the French National Research Agency (ANR) through the TEMPOS-NanoMax Equipex project. Zheng Fan thanks the Chinese Scholarship Council and FX-conseil for funding his PhD and Dr. P. Roura, Dr. F. Glas, Dr. F. Fortuna for fruitful discussion.